# Distributed actuator deformable mirror

**S.Bonora***

*CNR-IFN, Laboratory for Ultraviolet and X-ray Optical Research*

*via Trasea 7, 35131 Padova, Italy*

*Adaptica srl,*

*via Tommaseo 77, 35131 Padova, Italy*

*\*Corresponding author: bonox@dei.unipd.it*

## Abstract

In this paper we present a Deformable Mirror (DM) based on the continuous voltage distribution over a resistive layer. This DM can correct the low order aberrations (defocus, astigmatism, coma and spherical aberration) using three electrodes with nine contacts leading to an ideal device for sensorless applications.

We present a mathematical description of the mirror, a comparison between the simulations and the experimental results. In order to demonstrate the effectiveness of the device we compared its performance with the one of a multiactuator DM of similar properties in the correction of an aberration statistics. At the end of the paper an example of sensorless correction is shown.

Adaptive Optics (AO) technology is nowadays present in a lot of remarkable experiments [1,2]. It is a fundamental tool for large telescopes [3] and it is routinely used in extremely high power laser facilities [4,5]. Nonetheless AO is not wide spread diffused technology and its use is limited to touch and go experiments. There are several types of deformable mirrors on the market [6,7]. The most popular are membrane mirrors, which have obtained their diffusion thanks to their cheap price [8] but their use is limited by the low damage threshold. For high power laser applications bimorph mirrors are the most common choice because they can be easily associated to heat sinks [9]. Recently strong improvement in Adaptive Optics (AO) devices have been carried out by the introduction of magnetic and MEMS deformable mirrors which have respectively a very large stroke (up to 35µm) and resolution (up to about thousands actuators). All those technologies use a discrete series of actuators which are connected to a series of high voltage or current amplifiers in order to deform the mirror surface resulting in the need of complex and large electronic devices and cables.

All those systems are, in the most of the applications, actuated by the calculation of the desired shape using the influence functions matrix [6-9]. Hence in order to deform the mirror a preliminary knowledge of the deformation given by each actuator must be acquired with a wavefront sensor (WFs). In some cases, such as for example in scientific experiments, this is not possible thus optimization algorithms are successfully used [11].

Several studies have been carried out in order to reduce the complexity of adaptive optics devices trying to limit the number of actuators or to avoid the use of WFs in order to wide spread the use of adaptive devices [12]. Practical examples of applications, such as ophthalmology [14,15], astronomy [7,9] and scientific experiments [11], show that in the most of the cases the 90% of the aberrations weight is in the firsts three aberration orders [13]. Both [9,13] show that with a limited number of actuators it is possible to generate low order aberrations.

The device we present in this Letter addresses two problems: the former is the use of least actuators as possible, the latter is the direct generation the most important aberrations in a continuous actuators arrangement leading to an ideal system for open loop controls without the addition of a wavefront sensor. The main feature of of the DA-DM is that the actuators response is directly related to the optical aberrations allowing for a more versatile and straightforward use than conventional discrete actuators deformable mirrors. We will show that the voltages necessary to drive the mirror can be directly computed from the Zernike decomposition terms of the target.

In the paper we will describe the details of the device, we will present a mathematical description of the DA-DM and we will compare it with the measurement in the generation of the tilt, astigmatism, coma, defocus and spherical aberration. Then, in order to demonstrate the effectiveness of this device as adaptive optics corrector we will compare the performances of the DA-DM with a state of the art similar device and, at last, we show the use of the DA-DM in a sensorless application.

The device is a membrane electrostatic deformable mirror composed by a silvered nitrocellulose membrane 5μm thick. The membrane is suspended 70μm over the actuators by some spacers (Fig. 1). The geometry of the prototype we realized is: membrane diameter 19mm, active region 10mm, maximum voltage -130 / +130V. The prototype we used presented a little astigmatism (0.1μm peak to valley in the active region) in the rest position, at the maximum voltage the stroke of the DM was about 8μm. The deformable mirror was driven by a multichannel electronic driver for deformable mirrors (Adaptica IO32) which can supply 260V over 32 channels. We used nine channels connected to the contacts of the deformable mirror, the membrane was connected to a voltage reference of 130V. The DM is actuated [8] by the electrostatic pressure *p(x,y)* between the actuators and the metalized membrane which deforms the mirror *M(x,y)* according to the Poisson equation:

$$\Delta M(X,y) = \frac{1}{T} p(x,y) \tag{1}$$

In order to generate the low order aberration without any preliminary characterization of the mirror the device is composed by 3 actuators placed on three concentric rings. The first and the second actuators have 4 contacts each. The third one has just one contact (Fig. 1). The actuators are composed by a graphite layer 35μm thick which presents a sheet resistance of 1MOhm/sq. which continuously distributes the voltage. The current estimated for each channel is about 60μA with a power consumption for each actuator in the worst case condition of about 10mW.

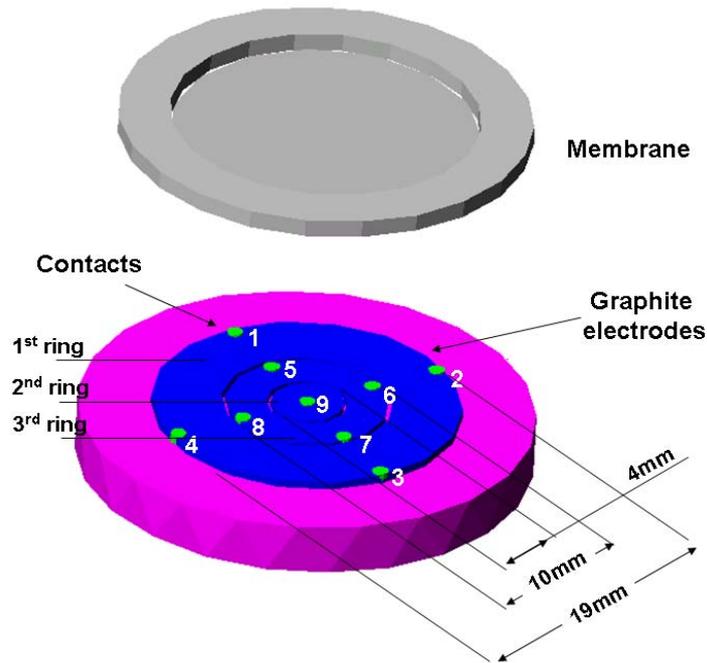

Fig. 1: diagram of the Distributed Actuator Deformable Mirror which shows the electrodes contacts and the resistive layers layout.

The voltage distribution over the resistive layer with resistivity $\rho$ can be computed again solving the Poisson equation for the scalar electric potential $U$:

$$\frac{1}{\rho}\Delta U(x,y) = 0 \qquad (2)$$

By a proper design of the deformable mirror the first actuator can be used for the generation of piston, tilt and astigmatism, the second actuator can be used for the generation of the coma and defocus, the third actuator is used for the generation of spherical aberration.

In order to solve both (1) and (2) we applied recursive technique on an grid domain. Some examples of voltage distribution are illustrated in Fig. 2.

The simulations show how the voltage of a single actuator is related to the mirror shape. Ref [13] explains that outside the active area are necessary *2N+1* actuators for the generation of *N* aberration orders. Here we show that, exploiting that the electrostatic pressure is proportional to the square of the voltage, driving the mirror with both positive and negative voltages it is possible to use less actuators. Fig. 2 shows that the application of opposite sign voltage to adjacent contacts generate an area of zero voltage and pressure (see Fig. 2, *OV* and *OP* dotted lines) leading to the generation of the astigmatism with 2N contacts outside the active area.

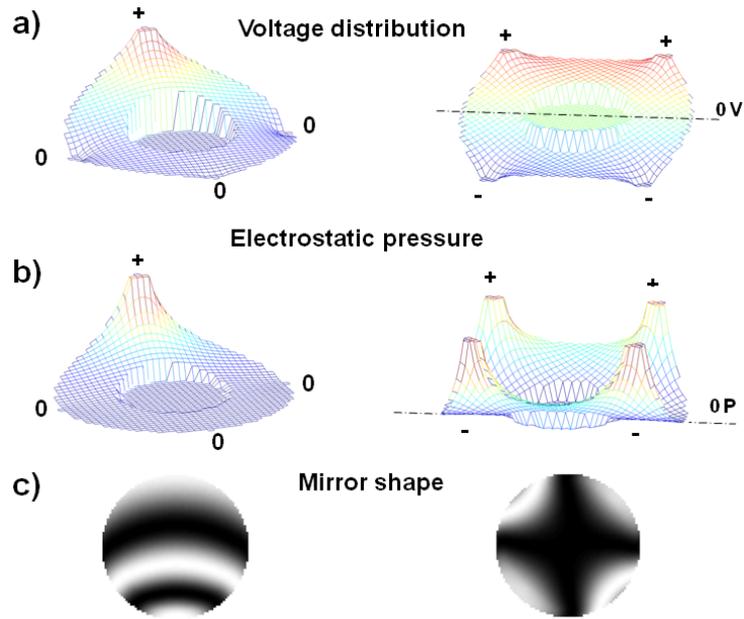

Fig. 2: simulation results. Left column: voltage distribution, electrostatic pressure and membranes shape obtained applying a voltage to one contact in the first electrode. Right column: example of generation of the astigmatism aberration. The voltage distribution uses both positive and negative voltages to generate a zero voltage distribution (dotted line 0V) in the middle of the membrane which corresponds to a zero electrostatic pressure (dotted line 0P).

In the following section we compare the simulation and measurement results for the generation of the tilt, defocus, coma and spherical aberrations. The measurements were carried out by interferometric technique of the deformable mirror surface (Zygo). The performance are evaluated computing the rms residual error with respect to target, the magnitude of the Zernike term and the spectral purity according to the formula:

$$P_i = \frac{Z_i}{\sqrt{\sum_{j=1}^{N} Z_j^2}} \qquad (3)$$

where $i$ is the index of the Zernike target aberration and $Z_j$ the Zernike spectrum of the measurements.

The tilt aberration can be generated applying to the first electrode an electrostatic pressure distributed over a tilted plane. This goal can be achieved applying the following voltages to the contacts 1-4, (see Fig. 1):

$$\begin{cases} V(1) = \sqrt{A_{tilt}} \frac{V_{max}}{2}[1+\sin(\alpha_t + \pi/2)] \\ V(2) = \sqrt{A_{tilt}} \frac{V_{max}}{2}[1+\sin(\alpha_t + \pi)] \\ V(3) = \sqrt{A_{tilt}} \frac{V_{max}}{2}[1+\sin(\alpha_t + 3\pi/2)] \\ V(4) = \sqrt{A_{tilt}} \frac{V_{max}}{2}[1+\sin(\alpha_t)] \end{cases} \quad (4)$$

Where $A_{tilt} = k_t \sqrt{Z_2^2 + Z_3^2}$ varies between (0-1) and controls the magnitude of the tilt while the angle is: $\alpha_t = \tan^{-1}(Z_3/Z_2)$. The simulations and measurement results show that the magnitude of the tilt is constant with the angle α with an average purity of 0.9963.

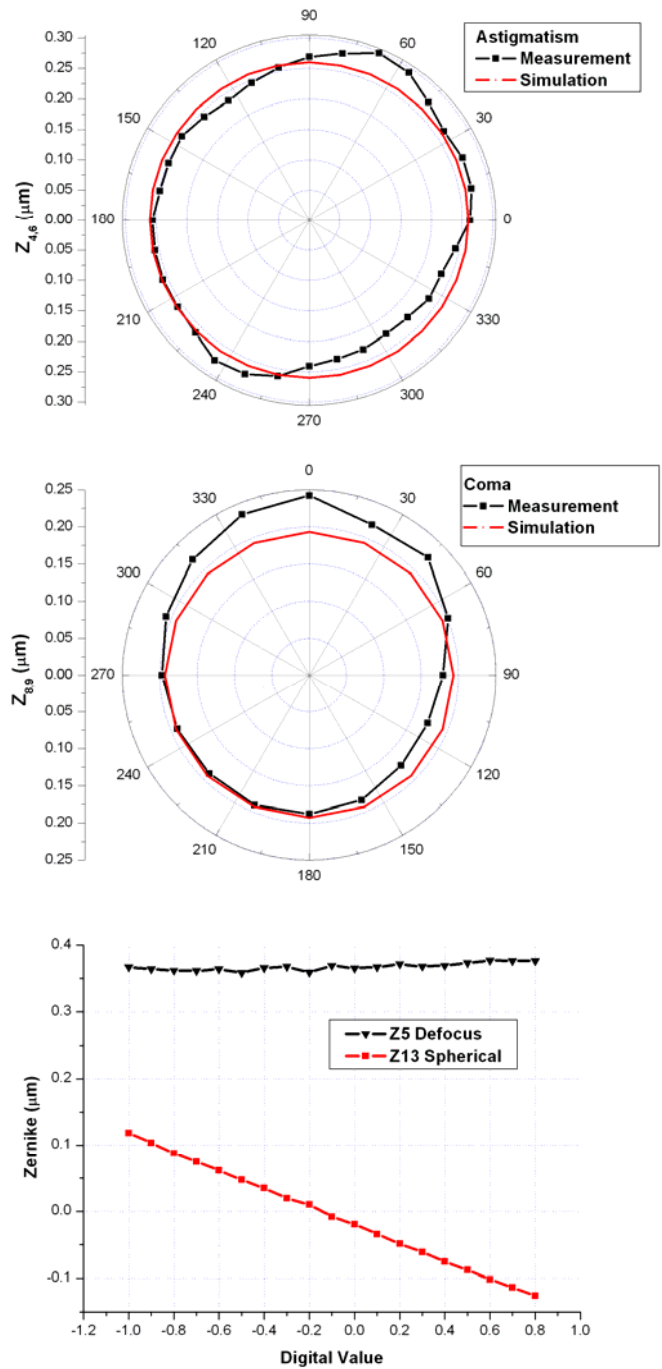

Fig. 3: measurements and simulation results comparison for astigmatism and coma aberration (Top and middle panel). Bottom panel: measurements of the demonstration of the generation of spherical aberration and defocus.

In order to generate astigmatism we follow the same rule which was valid for the tilt with the exception that using both positive and negative voltages yields to the generation of a symmetric electrostatic pressure (see for example Fig.2 second column).

Then the astigmatism aberration can be generated again on the first actuator by the following voltages:

$$\begin{cases} V(1) = k_l(\alpha_a)\sqrt{A_{ast}}V_{max}[\sin(\alpha_a + \pi/2)] \\ V(2) = k_l(\alpha_a)\sqrt{A_{ast}}V_{max}[\sin(\alpha_a + \pi)] \\ V(3) = k_l(\alpha_a)\sqrt{A_{ast}}V_{max}[\sin(\alpha_a + 3\pi/2)] \\ V(4) = k_l(\alpha_a)\sqrt{A_{ast}}V_{max}[\sin(\alpha_a)] \end{cases} \quad (5)$$

Where the parameter $A_{ast} = k_a\sqrt{Z_4^2 + Z_6^2}$ controls the magnitude ($k_a$ is a parameter that makes $A_{ast}$ between 0 and 1) and $\alpha_a = \tan^{-1}(Z_6/Z_4)$ and $k_l(\alpha_a) \propto 1/\max(V_{1-4})$ is a linearization factor. The simulations and measurements results are reported in Fig. 3 which shows the magnitude of the astigmatism aberration (in the Zernike expansion) in function of $\alpha_a$. The error in the generation of the astigmatism was measured by the spectral purity over the firsts five Zernike orders and has an average value over of 0.9840 with an average rms error on the target of 57nm.

Using the second actuator we have been able to generate coma aberration over a biased membrane[9] as reported in Fig 3 both theoretically and experimentally.

In this case the formulas were:

$$\begin{cases} V(5) = \sqrt{A_{coma}}V_{max}[\sin(\alpha_c + \pi/2)] \\ V(6) = \sqrt{A_{coma}}V_{max}[\sin(\alpha_c + \pi)] \\ V(7) = \sqrt{A_{coma}}V_{max}[\sin(\alpha_c + 3\pi/2)] \\ V(8) = \sqrt{A_{coma}}V_{max}[\sin(\alpha_c)] \end{cases} \quad (6)$$

where $A_{coma} = k_c\sqrt{Z_8^2 + Z_9^2}$, $k_c$ is a constant which makes $A_{coma}$ range between 0 and 1.

The voltages determinate by the formulas (6) generate a combination of coma and tilt which can be compensated using (4) in dynamic applications. The average spectral purity of the coma aberration is 0.92 with a mean deviation from the target of about 54nm.

Finally both defocus and spherical aberrations can be obtained using contacts 5-9. In order to generate those aberrations independently we used the following procedure. We computed the projection of the membrane shape generated by the actuator 2 ($S_2$, $V(5)=V(6)=V(7)=V(8)=V_{5-8}$) and the actuator 3 ($S_3$, $V(9)=V_9$) over the Zernike terms defocus $Z_5$ and Spherical Aberration $Z_{13}$.

$$\begin{cases} S_2 = a \bullet \hat{Z}_5 + b \bullet \hat{Z}_{13} \\ S_3 = c \bullet \hat{Z}_5 + d \bullet \hat{Z}_{13} \end{cases} \quad (7)$$

Since the Zernike modes are orthogonal the parameters $a, b, c, d$ can be computed by the scalar products:

$$a = S_2 \bullet \hat{Z}_5; b = S_2 \bullet \hat{Z}_{13}; c = S_3 \bullet \hat{Z}_5; a = S_3 \bullet \hat{Z}_{13}. \quad (8)$$

Now in order to generate independently combinations of spherical aberration and defocus the following parameters must be used:

$$\begin{cases} \hat{Z}_{13} = \dfrac{\frac{S_2}{a}}{(\frac{b}{a}+\frac{d}{c})} - \dfrac{\frac{S_3}{c}}{(\frac{b}{a}+\frac{d}{c})} = a_1 S_2 + b_1 S_3 \\ \hat{Z}_5 = \dfrac{\frac{S_3}{b}}{(\frac{a}{b}+\frac{c}{d})} - \dfrac{\frac{S_3}{d}}{(\frac{a}{b}+\frac{c}{d})} = c_1 S_2 + d_1 S_3 \end{cases} \rightarrow \begin{cases} V_{5-8} = V_{max}\sqrt{A_{sfe}a_1 + A_{def}c_1} \\ V_5 = V_{max}\sqrt{A_{sfe}b_1 + A_{def}d_1} \end{cases} \quad (9)$$

The coefficient $A_{sfe} = k_s Z_{13}$ and $A_{def} = k_d Z_5$ determines the magnitude of the spherical and defocus aberrations. In that measurements we kept fixed the defocus while the spherical aberration was linearly changed. Those values have been used for the measurements carried out in bottom panel of Fig. 3. We achieved a spectral purity of 0.9948 with an average rms error of 14nm.

Following from the previous section the knowledge of the DA-DM response is given by the measurements of the low order aberrations and not from the influence functions matrix. Thus the voltages which generates the aberration described by the Zernike coefficients $Z_1...Z_{13}$ is given by:

$$V_{1-9}=V_{defocus} + V_{Astigmatism} + V_{Coma} + V_{Spherical} =V_{5-9}(A_{def}) + V_{1-4}(A_{ast},\alpha_{ast})+ V_{5-8}(A_{coma},\alpha_{coma})+ V_{5-9}(A_{spher}) \quad (10)$$

where the complete description of the voltages can be found in 5-6, 9. Thus, using (10) it is possible to generate an aberration starting from the knowledge of its Zernike spectrum. In order to better understand the performances of the DA-DM with respect to a state of the art device, we compare its performances with the one of a commercially available 32 actuators DM (PAN deformable mirror, Adaptica srl).

We used the information obtained from the previous simulations to correct the statistic of an aberrated eye population [14] and we compared it with the correction of the PAN mirror by the pseudoinversion of the influence function matrix as explained in [16]. In order to well understand the DM correction we scaled the maximum stroke of the PAN mirror to find the one which allows to correct with the same average residual the eye population. The result is that the DA-DM is equivalent to a *Pan* DM with a stroke reduced of 0.75. This can be explained observing that (see Fig. 2) between adjacent contacts the voltage drops generating less electrostatic pressure than the one obtained forcing it with discrete actuators.

The last test for the demonstration of the advantages of this technology was the correction of the quality of an optical system 15X build with just spherical elements. The correction procedure was carried out manually by an operator optimizing the sharpness of the test image acting directly in sequence on the defocus, astigmatism, spherical aberration and coma. We have been able to improve the resolution of the image by the maximization of the visibility of the test target lines of the group 7, element 6 (228 lines pair/mm). Fig. 4 shows the starting image, the optimization during the first iteration and the final image. During a three iterations manual process we were able to achieve nearly full visibility.

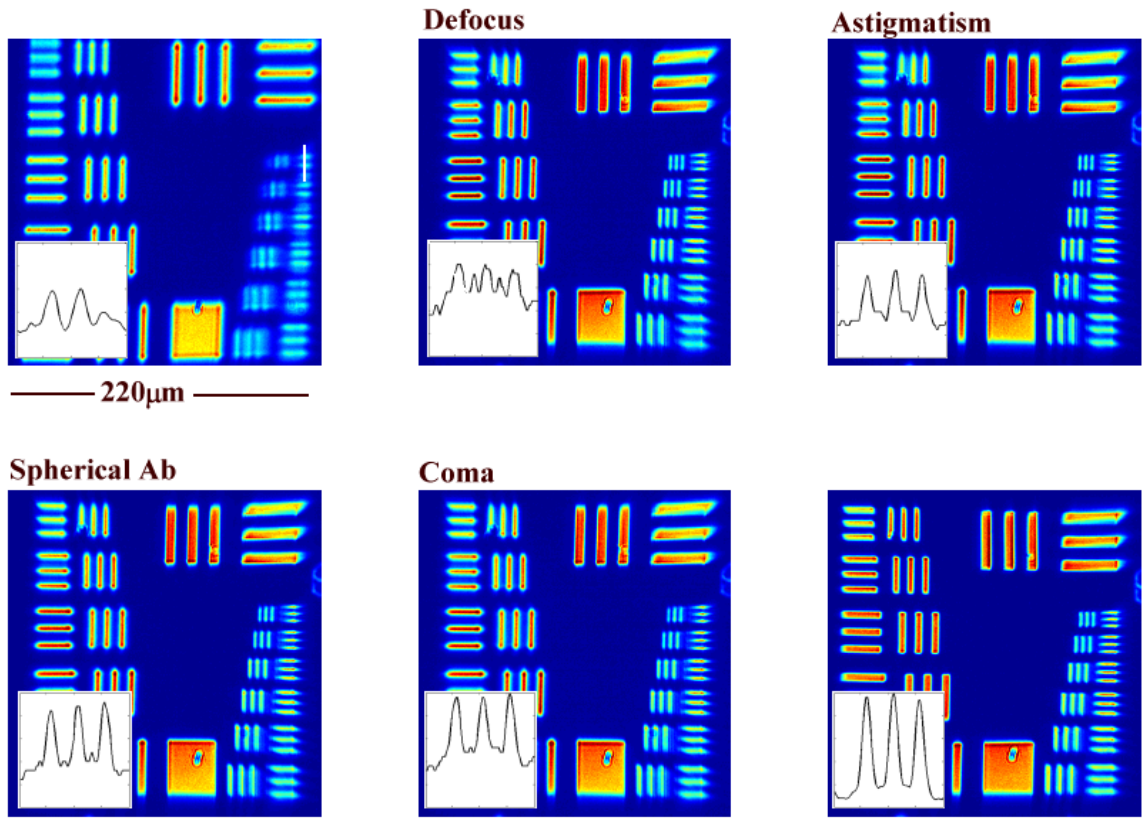

Fig. 4: manual correction of a test image in a 15X optical system. The first panel reports the initial image. Then panel 2-5 reports the image after the correction of the defocus, astigmatism, spherical aberration and coma in the first iteration. Then the last panel shows the image after a three iterations. The insets reports the cross section of the group 7, element 6 (228lines pair/mm).

In conclusion we have shown that the technique of the distributed actuator can be used for the realization of non pixellated deformable mirrors introducing a direct control of low order aberrations with an excellent quality and linearity. The electrostatic DA-DM have been compared to a commercial 32 actuators DM demonstrating to have nearly the same performances in the correction of an eye aberration statistic. Moreover the DA-DM technique allows to directly connect the effect of the actuators to the optical quality through the generation of Zernike polynomials. This property was used in a test setup for a demonstration of sensorless correction. The concept of this design, which exploits resistive layers, can be implemented in other technologies such as bimorph mirrors.

**Acknowledgments.**

The realization of this devices was funded by Adaptica srl (Patent pending M3101733/IT). The measurements were carried out at the CNR-INFM LUXOR Lab. The author is grateful to Cosmo Trestino for the device design and to Fabio Frassetto and Tommaso Occhipinti for the useful discussions.